\documentclass[namedreferences,hyperref,optionalrh]{spr-sola}
\usepackage{graphicx}        
\usepackage{amsmath}         
\usepackage{color}           
\usepackage[mathlines]{lineno}

\renewcommand{\vec}[1]{{\mathbfit #1}}

\chardef\us=`\_

\begin{document}

\begin{frontmatter}
\title{The two-stream instability generation around Moon: Effect of Interplanetary Magnetic Field during Solar Wind - Lunar Plasma Interaction}

\author[addressref={aff1},corref,email={vipin\_ky@vssc.gov.in}]{\inits{Vipin K.}\fnm{Vipin K.}~\snm{Yadav}\orcid{0000-0002-1470-8443}}

\author[addressref={aff2}]{\fnm{Abhinav}~\snm{Singh}}

\author[addressref={aff3}]{\fnm{Rajneesh}~\snm{Kumar}\orcid{0000-0002-1854-4589}}


\address[id=aff1]{Space Physics Laboratory (SPL), Vikram Sarabhai
Space Centre (VSSC), Thiruvananthapuram 695022, India}

\address[id=aff2]{Aryabhatta Research Institute of Observational Sciences (ARIES), Nainital 263001, India}

\address[id=aff3]{Department of Physics, Banaras Hindu University (BHU), Varanasi 221005, India}

\runningauthor{Yadav et al.}
\runningtitle{\textit{Solar Physics} Lunar TSI: IMF}


\begin{abstract}
The relative motion of two interpenetrating streams of charged particles usually leads to the generation of two-stream instability (TSI) and eventually result in the onset of non-linear plasma processes such as the turbulence or the plasma waves. A natural example of such an event is the solar wind interaction with the lunar electron plasma where the interplanetary magnetic field (IMF) is embedded within the solar wind. The inclusion of IMF in the solar wind – lunar plasma interaction modifies the dispersion relation of the TSI and an angular cyclotron frequency term appears in the denominator of the leading term and hence leads to the change in the parameters such as the instability growth rate as now it depends on the solar wind electron velocity, solar wind and the lunar electron plasma density, and the IMF magnitude.  It is observed that the growth rate increases fast with the increase in the magnetic field initially but the increase slows down on further increasing the magnetic field thereby smoothening the top. From the particle-in-cell (PIC) simulations, it is observed that during the solar wind IMF interaction with lunar plasma, the non-energetic background electrons make a shield around the solar wind electrons in the vortices formed. It is further observed that those lunar electrons which are not participating either in the vortices formation or in the shielding of solar wind electrons, start moving in the direction of the incoming solar wind. These observations indicate this interaction is capable in getting converted into a non-linear physical process in the lunar plasma environment.
\end{abstract}
\end{frontmatter}


\section{Introduction}
     \label{S-Introduction} 

The Sun is a persistent source of energy and plasma, constantly emitting electromagnetic radiation and a supersonic stream of charged particles, known as the solar wind, into the heliosphere. This outflow carries the interplanetary magnetic field (IMF), which originates in the solar corona and is frozen into the outward-flowing plasma due to its high electrical conductivity. As the solar wind moves through the interplanetary medium (IPM), the IMF becomes wrapped into an Archimedean spiral shape, caused by solar rotation combined with radial plasma outflow \citep{Schwenn1990}. Beyond the acceleration zone, the solar wind typically reaches a steady terminal velocity, while its particle density decreases roughly as $1/r^2$, where $r$ is the heliocentric distance. In this regime, plasma pressure and magnetic pressure are comparable (plasma beta, $\beta \approx 1$) \citep{Russell2001}, greatly influencing plasma dynamics in the IPM and controlling the transport of energetic particles, turbulence development, and interactions with planetary magnetospheres and ionospheres \citep{Kivelson1995}.

For decades, the Moon was regarded as a passive object in the solar wind due to its lack of a magnetic field and the weak, surface-bound nature of its exosphere \citep{Stern1999}. This paradigm shifted significantly following dedicated plasma and particle measurements from a series of lunar missions. These include \textit{Chandrayaan-1} by India, \textit{Kaguya} by Japan, \textit{Chang’E-1} by China, and NASA’s \textit{Lunar Reconnaissance Orbiter} (LRO), and the dual-probe \textit{ARTEMIS} mission (Acceleration, Reconnection, Turbulence and Electrodynamics of the Moon’s Interaction with the Sun). Observations from these missions have revealed a complex suite of interaction phenomena. These include the sputtering of atoms from the regolith by solar wind ions, backscattering of solar wind protons as ENAs, formation of localized mini-magnetospheres over crustal magnetic anomaly regions, pick-up ion formation from surface-derived neutrals, and the existence of multiple coexisting plasma populations in the near-lunar environment \citep{Barabash2012, Bhardwaj2015}.

The lunar plasma environment has been extensively investigated using radio occultation (RO) techniques \citep{Choudhary2020} which offer sensitive diagnostics of electron density profiles near planetary bodies. RO measurements reported by \citet{Choudhary2016} indicated that near the lunar surface, electron densities typically range from $300 ~cm^{-3}$ to $1000 ~{cm}^{-3}$, with a rapid decline in altitude and a near-complete depletion beyond $\sim 40 ~km$. These results support the presence of a highly surface-bound plasma layer, consistent with expectations for a collisionless and unmagnetized body. In scenarios where the dynamic interaction with the solar wind is temporarily neglected, local enhancements in surface electron density can reach as high as $3 \times 10^4 ~{cm}^{-3}$, with dominant ion species such as Ar$^+$, Ne$^+$, and He$^+$ derived from surface release processes and solar wind implantation \citep{Ambili2022}. However, under typical solar wind conditions, the absence of an intrinsic magnetic field leads to efficient scavenging of positive ions by the solar wind, resulting in a tenuous residual ionosphere with densities often below $5 ~{cm}^{-3}$ \citep{Choudhary2016}.

In addition to large-scale flow and kinetic processes, the lunar environment also support plasma wave activity and instabilities arising from streaming populations \cite{Yadav2020, Tolba2024}. Streaming plasma instabilities, such as the two-stream instability (TSI), are predicted to exist in space such as the interstellar medium (ISM) where fast-moving cosmic rays propagate through a background plasma, driving wave growth through velocity-space anisotropy \citep{Yadav2018}. Similar mechanisms are expected in the lunar wake and near-surface regions, where reflected or pick-up ions interact with ambient solar wind plasma. Indeed, observations from the \textit{ARTEMIS} mission have confirmed the presence of electrostatic and electromagnetic plasma waves around the Moon, including Langmuir waves, ion-acoustic waves, and whistler-mode emissions \citep{Yadav2020}. These wave activities are closely linked to the generation of turbulence, energy dissipation, and plasma transport processes, thereby contributing to the broader understanding of plasma dynamics in weak-field, atmosphere-less planetary environments \citep{Temerin1989}.

The lunar crustal magnetic field interaction with the solar wind electrons is observed with Lunar Prospector and the enhanced magnetic field along with the increase in the low energy electron fluxes is observed in the solar wind over lunar surfaces where the crustal magnetic field is strong. The electron energization, in some cases, led to the triggering of low frequency plasma wave appearance \citep{Halekas2008}. Also, several magnetic anomalies on the lunar surface are identified with measurements from a magnetometer onboard Lunar Prospector from a low altitude \citep{Hood2011}. The solar wind plasma interaction with the crustal magnetic field on the lunar surface is not the same as its interaction with planetary magnetosphere as the lunar magnetic field spatial scale is much less than the radius of Moon. The solar wind – lunar crustal magnetic field interaction is studied using a 3-dimensional self-consistent hybrid plasma model and found that the crustal magnetic field shield the lunar surface from the incoming solar wind plasma in a local region and gives rise to an electrostatic potential above from the surface of Moon \citep{Fatemi2015}. Recently, the solar wind and interplanetary magnetic field interaction with the local lunar crustal magnetic field is also studied using measurements by the Acceleration, Reconnection, Turbulence and Electrodynamics of Moon's Interaction with the Sun (ARTEMIS) spacecraft. In this analysis it is observed that there is very little impact of the lunar crustal magnetic field support the generation of very low frequency plasma waves and the IMF affected solar wind electrons acquire a drift which is normal to the lunar magnetic field triggering the flow of Hall current which can change the magnetic field strength near the lunar surface \citep{Liu2024}. However, in this study the effect of localized lunar crustal field is not considered and shall be accounted for in another study in the near future.

The present study builds upon this growing interest in observational and theoretical work to further investigate the plasma characteristics in the vicinity of the Moon thereby modifying the classical dispersion relation of the TSI for a magnetized solar wind; obtaining the updated TSI growth factor/rate; and the parameters influencing the generation of the TSI in lunar plasma environment. The other scientific objective is to visualize the TSI generation employing the particle-in-cell (PIC) simulations for various IMF magnitudes, thereby advancing our understanding of the solar wind interaction with airless, unmagnetized bodies and the associated implications for surface–plasma coupling, exospheric structure, and heliophysical plasma processes. These studies can lead to a better design for future lunar landing and sample return missions. 

The paper is organized as follows: The magnetized solar wind and the solar wind - lunar plasma interaction is described in Section 2; Section 3 presents the streaming plasma instabilities including the earlier worked carried out in this regard; In Section 4, the modified dispersion relation of the TSI is obtained analytically; Section 5 is dedicated to the analysis of lunar TSI with magnetized solar wind; The PIC simulations are carried out in Section 6. Section 7 discusses the analytical and numerical observations of this work and the conclusions drawn from this scientific study are compiled in Section 8.

\section{Solar Wind - Lunar Plasma Interaction}

Plasma emitted from the solar corona flows outward as a supersonic, collisionless stream - the solar wind, which permeates the heliosphere and governs many space weather interactions. This flow consists primarily of electrons, protons, alpha particles (He$^{2+}$), and minor contributions from heavier ions. Near 1~AU, the solar wind typically maintains a number density around $8 ~{cm}^{-3}$ and an average speed of $440 ~{km/s}$, though both quantities fluctuate considerably due to the dynamic solar environment \citep{Gosling2007}.

When the solar wind encounters a planetary body, the nature of the interaction depends on whether there is an intrinsic magnetic field and a substantial atmosphere on that body. For magnetized planets such as Earth, the solar wind is deflected by the planetary magnetic field, creating a well-defined bow shock and magnetosphere. Conversely, for airless, unmagnetized bodies such as the Moon, the interaction happens directly near the surface or within its tenuous plasma environment. The Moon, on the other hand, lacks both the global magnetic field and a dense atmosphere, resulting in a unique plasma interaction characterized by direct solar wind impact, local field-aligned flow, and surface-bound exospheric processes \citep{Luhmann1995, Barabash2009}. The Moon has a very thin ionosphere, extending from lunar surface to 100 km which gets generated due to the ionization of the atmospheric neutrals by the ultraviolet photons of the solar radiation \citep{Stern1999}. Despite of it being tenuous, the lunar plasma density in lunar ionosphere is variable \citep{Vyshlov1976}; \citep{Withers2021} due to the variability in solar conditions \citep{Freeman1976}.

The unmagnetized Moon, having only a tenuous exosphere, allows the solar wind plasma to interact almost unhindered with its surface which induces the surface currents and electric fields governed by Faraday’s law, resulting in fundamentally different coupling processes \citep{Barabash2012}. These interactions become observable primarily when the Moon moves outside Earth's magnetotail and magnetosheath—conditions that occur for about ten days on either side of the new Moon phase, depending on the position of Earth's bow shock \citep{Ness1972}.

In the case of Moon, the IMF lines frozen in the solar wind, remain largely unaltered and penetrate the near-lunar environment, leading to a flow symmetry about a plane defined by the solar wind velocity vector and the IMF direction, passing through the lunar center. When the solar wind interacts with the lunar surface, ions are neutralized and re-emitted as energetic neutral atoms (ENAs), contributing to surface sputtering and the exospheric composition \citep{Whang1968, Dhanya2013, Bhardwaj2015}.

Direct observational evidence of these interactions came from the Sub-keV Atom Reflecting Analyzer (SARA) payload onboard India’s Chandrayaan-1 mission. SARA detected sputtered surface atoms and backscattered neutral hydrogen produced by solar wind protons impacting the lunar regolith. These measurements confirmed that the Moon’s interaction with the solar wind leads to substantial surface modification and particle release into the surrounding space environment \citep{Barabash2009}.

\section{Streaming Instabilities in Lunar Plasma}

Within the solar wind, electrons are divided into two populations: a dense, cooler core and a tenuous, hotter halo. Both groups have thermal velocities that exceed the bulk flow speed of the solar wind. Along the IMF direction, electrons drift more rapidly than ions, whereas in the perpendicular direction, both species exhibit nearly the same drift speed \citep{Russell2001}. These kinetic characteristics influence how the solar wind interacts with objects immersed in it. The solar wind interaction with the plasma environment of the planetary bodies gives rise to complex physical phenomena including plasm instabilities such as Rayleigh–Taylor (RT) instability in which a less dense plasma supports a higher density plasma subjected to a gravitational force; Kelvin–Helmholtz (K-H) instability where a velocity shear is present in various plasma layers; Buneman instability where a plasma beam passes through a plasma, etc. \citep{Chen2016}.

Plasma instabilities especially the streaming plasma instabilities are predicted to be present in regions of space such as the inter-stellar medium (ISM) where the high energy cosmic rays protons propagate through the interstellar plasma thereby invoking a two-stream instability \citep{Yadav2018}. The result of these instabilities is often in the form of generation of plasma waves, a number of plasma waves are observed around the Moon also \citep{Yadav2020}.

The two-stream instability (TSI) is a tiny perturbation in the electron plasma density, due to which an electric field grows with time and this growth is facilitated by the electron plasma in the background \citep{Chen2016}. On Moon, the TSI develops as incoming solar wind electrons interact directly with the tenuous lunar exospheric electron population, where conditions favor electrostatic instabilities. For lunar TSI, the growth factor, the threshold condition for TSI to get triggered and the solar wind and lunar plasma parameter contribution to trigger TSI is investigated along with the effect of these parameters in the evolution of TSI \citep{Chakraborty2022, Chakraborty2023}.

The particle-in-cell (PIC) \citep{Mocz2020} simulations are employed to visualize the evolution of TSI in phase space. These simulations show mixing of solar wind and lunar electrons, accompanied by the formation of periodic, closed vortices—signatures of nonlinear wave–particle coupling. The instability leads to electron bunching, which enhances the local electron number density and may explain observed enhancements in the lunar ionosphere  \citep{Chakraborty2023}.

To further investigate this behavior, this analysis is extended by accounting an energetic (hot) electron population also from the lunar plasma environment. During the analysis, the hot electron fraction is varied from 1\% to 25\% of the total electron population. These hot electrons significantly modified the TSI dispersion relation, lowered the instability threshold, and accelerated its growth \citep{Yadav2024a, Yadav2024b}.

The simulations incorporating 'hot' electrons demonstrated faster onset of the electron bunching and more rapid growth of the instability. As the hot electron fraction increased, the time required for instability onset decreased. These results suggest that even a modest energetic population can strongly influence the development of TSI and may contribute to spatial and temporal variability in the lunar near-surface plasma environment \citep{Yadav2024a, Yadav2024b}.

In studies, so far, the conditions necessary for TSI to develop in the lunar environment are analytically examined, the associated growth rates/factors are computed, and the threshold parameters are identified. These results show that the instability onset depends on the beam (solar wind) density and velocity, as well as the ambient lunar electron density. The dispersion relation reveals a range of wave numbers ($k$) that can support unstable modes, although only a few exhibit significant growth at each altitude. Each location supports one dominant, fastest-growing mode, while others remain stable or decay.
 
\section{The Modified Dispersion Relation of TSI} 

The beam plasma is considered with following assumptions:

\begin{itemize}

\item Plasma is infinite in extent, is homogeneous in density and is isotropic with respect to all plasma parameter.

\item Plasma considered to be cold in nature, so neglecting thermal effect.

\item Here, magnetic field is considered, so this plasma is magnetized plasma.

\item The oscillation considered will be longitudinal in nature.

\item The ions do not move and the mobility is due to electrons only.

\item Beam density will be much lower than plasma density.

\end{itemize}

\noindent
The plasma parameters taken are the plasma density, $n_e$ and the electron velocity, $v_e$ and the beam plasma parameters are the beam electron density, $n_b$ and the electron velocity, $v_b$.

The perturbation in the beam plasma parameters is given as

\begin{equation}
 n_b = n_{0b} + \tilde{n}_b \\ \hspace*{20mm}
 v_b = v_{0b} + \tilde{v}_b
\end{equation}

The equation of motion for beam plasma is given as

\begin{equation}
 mn_b\left[\frac{\partial \vec{v_b}}{\partial t} + \vec{v_b}.\nabla\vec{v_b} \right] = -en_b[\vec{E} + (\vec{v} \times \vec{B})]
\end{equation}

and the continuity equation is given as

\begin{equation}
 \frac{\partial n_b}{\partial t} + \nabla.(n_b\vec{v_b}) = 0
\end{equation}

In this case only longitudinal waves with $k \parallel \vec{E}$. are considered, hence $k$ and $\vec{E}$ are along the x-axis. Also, $\vec{B}$ is along z-axis. Therefore,

\begin{equation}
 \vec{k} = k\hat{x}; \hspace*{10mm} \vec{E} = E\hat{x}; \hspace*{10mm} \vec{B} = B\hat{z}
\end{equation}

The equations (2) and (3) are normalized using equation (1). Hence, equation (2) becomes

\begin{equation}
 \frac{\partial \tilde{\vec{v_b}}}{\partial t} + \vec{v_{0b}}.\nabla \tilde{\vec{v}}_b = - \frac{e}{m}[\vec{E} + (\tilde{\vec{v}}_b \times \vec{B})]
\end{equation}

and equation (3) becomes

\begin{equation}
 \frac{\partial \tilde{n}_b}{\partial t} + \nabla.(n_{0b}\tilde{\vec{v}}_b) + \nabla.(n_b\tilde{\vec{v}}_{0b}) = 0
\end{equation}

The wave like solution is considered such as

\begin{equation}
 n = n_{0b}e^{i(kx - \omega t)}
\end{equation}

which gives

\begin{equation}
 \nabla = ik,  \hspace*{20mm} \frac{\partial}{\partial t} = -i\omega
\end{equation}

Using values in equation (8), equation (5) becomes

\begin{equation}
 -i\omega\tilde{v}_b + ikv_{0b}\tilde{v}_b = - \frac{e}{m}[E + \tilde{v}_bB]
\end{equation}

Equation (9) is resolved into x-component as

\begin{equation}
 -i\omega\tilde{v}_{b(x)} + ikv_{0b(x)}\tilde{v}_{b(x)} = -\frac{e}{m}[E + \tilde{v}_{b(y)}{B_0}]
\end{equation}

The y-component is given as

\begin{equation}
 -i\omega\tilde{v}_{b(y)} + ikv_{0b(y)}\tilde{v}_{b(y)} = -\frac{e}{m}\tilde{v}_{b(x)}{B_0}
\end{equation}

which becomes

\begin{equation}
 \tilde{v}_{b(y)} = \frac{-i(e/m)B_0}{(\omega - kv_{0b(y)})}\tilde{v}_{b(x)}
\end{equation}

And the z-component is given as

\begin{equation}
 -i\omega\tilde{v}_{b(z)} + ikv_{0b(z)}\tilde{v}_{b(z)} = 0
\end{equation}

Again using values in equation (8), equation (6) becomes

\begin{equation}
 -i\omega \tilde{n}_b + ik\tilde{v}_bn_{0b} + ik\tilde{n}_bv_{0b} = 0
\end{equation}

which can also be written as

\begin{equation}
 \tilde{n}_b = \frac{k\tilde{v}_b}{(\omega - kv_{0b})}n_{0b}
\end{equation}

Inserting equation (12) in equation (10), we get

\begin{equation}
 -i\omega\tilde{v}_{b(x)} + ikv_{0b(x)}\tilde{v}_{b(x)} = -\frac{e}{m}[E + \frac{-i(e/m)B_0}{(\omega - kv_{0b(y)})}\tilde{v}_{b(x)}{B_0}]
\end{equation}

On rearranging, it becomes

\begin{equation}
 \tilde{v}_{b(x)} = \frac{-i(e/m)E}{\left[(\omega - kv_{0b(x)}) + (\frac{eB_0}{m})^2 \frac{1}{(\omega - kv_{0b(y)})}\right]}
\end{equation}

Now, inserting equation (17) in equation (15) gives

\begin{equation}
 \tilde{n}_b = \frac{\frac{-i(e/m)E}{\left[(\omega - kv_{0b(x)}) + (\frac{eB_0}{m})^2 \frac{1}{(\omega - kv_{0b(y)})}\right]}}{(\omega - kv_{0b(x)})}n_{0b}k
\end{equation}

or

\begin{equation}
 \tilde{n}_b = \frac{-ikE}{(\omega - kv_{0b(x)})\left[(\omega - kv_{0b(x)}) + (\frac{eB_0}{m})^2 \frac{1}{(\omega - kv_{0b(y)})}\right]}(\frac{en_{0b}}{m})
\end{equation}

Now, the plasma system is considered for which also the equation of motion and the equation of continuity is employed which is as follows

The equation of motion gives

\begin{equation}
 mn_e\left[\frac{\partial \vec{v_e}}{\partial t} + (\vec{v_e}.\nabla\vec{v_e}) \right] = -en_e\vec{E}
\end{equation}

and the continuity equation is written as

\begin{equation}
 \frac{\partial n_e}{\partial t} + \nabla.(n_e\vec{v_e}) = 0
\end{equation}

The linearization is carried out as

\begin{equation}
 n_e = n_{0e} + \tilde{n}_e \\ \hspace*{20mm}
 \vec{v}_e = \vec{v}_{0e} + \tilde{\vec{v}}_e
\end{equation}

Which gives the equation of motion as

\begin{equation}
 m\left[\frac{\partial (\vec{v}_{0e} + \tilde{\vec{v}}_e)}{\partial t} + [(\vec{v}_{0e} + \tilde{\vec{v}}_e)\nabla.(\vec{v}_{0e} + \tilde{\vec{v}}_e)] \right] = -e\vec{E}
\end{equation}

which reduces to

\begin{equation}
 \frac{\partial (\tilde{v}_e)}{\partial t} = -\frac{e}{m}E
\end{equation}

The continuity equation from equation (21) becomes

\begin{equation}
 \frac{\partial (n_{0e} + \tilde{n}_e)}{\partial t} + \nabla.(n_{0e} + \tilde{n}_e)(\vec{v}_{0e} + \tilde{\vec{v}}_e) = 0
\end{equation}

and which reduces to

\begin{equation}
 \frac{\partial (\tilde{n}_e)}{\partial t} + \nabla.(n_{0e})(\tilde{\vec{v}}_e) = 0
\end{equation}

Again, on invoking a wave like solution here equation (24) becomes

\begin{equation}
 -i\omega\tilde{v}_e = -\frac{e}{m}E
\end{equation}

which becomes

\begin{equation}
 \tilde{v}_e = -i\frac{(e/m)}{\omega}E
\end{equation}

And equation (26) becomes

\begin{equation}
 -i\omega\tilde{n}_e + ikn_{0e}\tilde{v}_e = 0
\end{equation}

\noindent
or

\begin{equation}
 \tilde{n}_e = \frac{k\tilde{v}_e}{\omega}n_{0e}
\end{equation}

\noindent
Inserting equation (28) here, we get

\begin{equation}
 \tilde{n}_e = \frac{-ik\frac{(e/m)}{\omega}E\tilde{v}_e}{\omega}n_{0e} = \frac{-ikE}{\omega^2}(\frac{en_{0e}}{m})
\end{equation}

\noindent
Now Gauss law gives

\begin{equation}
 \nabla.\vec{E} = \frac{\rho}{\epsilon_0} = \frac{\rho_b}{\epsilon_0} + \frac{\rho_f}{\epsilon_0} = \frac{e[n_i - n_e]}{\epsilon_0} + \frac{\rho_f}{\epsilon_0}
\end{equation}

\noindent
here $\rho$ is the total and $\rho_b$ and $\rho_f$ are the bound and free charge density respectively.

\noindent
Applying linearization, equation (32) becomes

\begin{equation}
 \nabla.\vec{E} = \frac{e}{\epsilon_0}[n_{ib} - n_{0b} - \tilde{n}_b + n_{ie} - n_{0e} - \tilde{n}_e] + \frac{\rho_f}{\epsilon_0}
\end{equation}

\noindent
According to plasma quasi-neutrality

\begin{equation}
 n_{ib} = n_{0b}; \hspace*{20mm} n_{ie} = n_{0e}
\end{equation}

\noindent
which reduces equation (33) to

\begin{equation}
 \nabla.\vec{E} = \frac{e}{\epsilon_0}[- \tilde{n}_b - \tilde{n}_e] + \frac{\rho_f}{\epsilon_0}
\end{equation}

\noindent
Substituting $\tilde{n}_b$ from equation (19) and $\tilde{n}_e$ from equation (31) in equation (35), we get

\begin{center}
\begin{eqnarray}
 \nabla.\vec{E} =
 - \frac{e}{\epsilon_0}[\frac{-ikE}{(\omega - kv_{0b(x)})\left[(\omega - kv_{0b(x)}) + (\frac{eB_0}{m})^2 \frac{1}{(\omega - kv_{0b(y)})}\right]}(\frac{en_{0b}}{m}) + \frac{-ikE}{\omega^2}(\frac{en_{0e}}{m})]
 \nonumber \\
  +  \frac{\rho_f}{\epsilon_0}
\end{eqnarray}
\end{center}

\noindent
On rearranging, it becomes

\begin{eqnarray}
 \nabla.\vec{E} \left[1 - [\frac{1}{(\omega - kv_{0b(x)})\left[(\omega - kv_{0b(x)}) + (\frac{eB_0}{m})^2 \frac{1}{(\omega - kv_{0b(y)})}\right]}\left(\frac{e^2n_{0b}}{m\epsilon_0}\right) + \frac{1}{\omega^2}\left(\frac{e^2n_{0e}}{m\epsilon_0}\right)]\right]
 \nonumber \\
 \hspace{-40mm} = \frac{\rho_f}{\epsilon_0}
 \label{divE1}
\end{eqnarray}

\noindent
Taking beam plasma angular frequency ($\omega_b$) and lunar plasma angular frequency ($\omega_p$) as  

\begin{equation}
 \omega_b^2 = \left(\frac{e^2n_{0b}}{m\epsilon_0}\right); \hspace*{20mm} \omega_p^2 = \left(\frac{e^2n_{0e}}{m\epsilon_0}\right)
\end{equation}

\noindent
Equation (\ref{divE1}) becomes

\begin{equation}
\nabla.\vec{E} \left[1 - \frac{\omega_b^2}{(\omega - kv_{0b(x)})\left[(\omega - kv_{0b(x)}) + (\frac{eB_0}{m})^2 \frac{1}{(\omega - kv_{0b(y)})}\right]} - \frac{\omega_p^2}{\omega^2}\right] = \frac{\rho_f}{\epsilon_0}
\label{divE2}
\end{equation}

\noindent
It is compared with the Gauss law for dielectrics as

\begin{equation}
 \nabla.\vec{D} = \rho_f; \hspace*{20mm} \nabla.\epsilon \vec{E} = \rho_f
 \label{GL}
\end{equation}

\noindent
Comparing equation (\ref{GL}) with equation (\ref{divE2}), we have

\begin{eqnarray}
\epsilon_0\left[1 - \frac{\omega_b^2}{(\omega - kv_{0b(x)})\left[(\omega - kv_{0b(x)}) + (\frac{eB_0}{m})^2 \frac{1}{(\omega - kv_{0b(y)})}\right]} - \frac{\omega_p^2}{\omega^2}\right]
\nonumber \\
= \epsilon = \epsilon(\omega, k)
\label{epsilon}
\end{eqnarray}

\noindent
For normal mode plasma oscillations, equation (\ref{epsilon}) = 0 and for $kv_{0b(x)} = kv_{0b(y)} = kv_{0b} = \omega_d$ (beam angular frequency), and $eB_0/m = \omega_c$  (beam cyclotron frequency) it becomes

\begin{equation}
1 = \frac{\omega_b^2}{(\omega - \omega_d)^2 + \omega_c^2} + \frac{\omega_p^2}{\omega^2}
 \label{DisRel}
\end{equation}

\noindent
This is the dispersion relation of the two-stream instability generated due to the interaction of solar wind along with IMF with the lunar electron plasma. For $\omega_c = 0$ when $B_0 = 0$, the dispersion relation is reduced again to its classical form.

\begin{figure}[h]
\centering
\includegraphics[height=3.0in,width=3.25in]{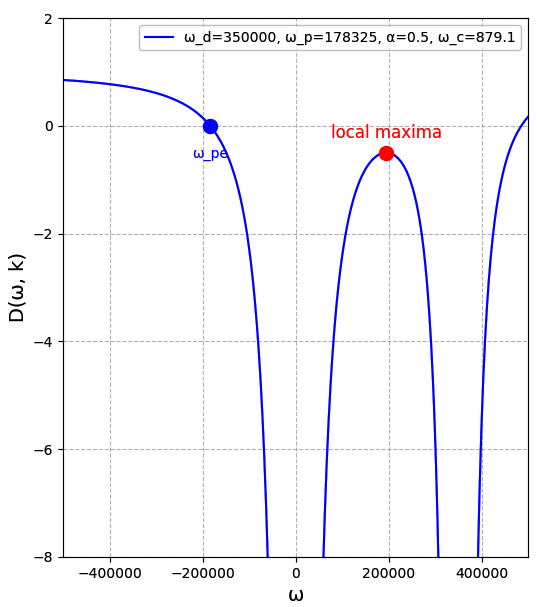}
\caption{The TSI dispersion relation.}
\label{DRP}
\end{figure}

\noindent
The dispersion relation for the TSI generated due to the interaction of the magnetized solar wind with lunar electrons is shown in Figure \ref{DRP}.

\noindent
Here, the electron plasma density is taken as $n_e = 10 ~cm^{-3}$ which gives $\omega_p = 178325 ~rad/s$, the IMF is taken as $B_0 = 5 ~nT$ \citep{Gosling2007} which gives $\omega_c = 879.1 ~rad/s$, $n_b = 5 ~cm^{-3}$ \citep{Gosling2007} which makes $\alpha = n_b/n_e = 0.5$ and for $k = 1$, $v_{0b} = 3.5 \times 10^5 ~m/s$ (typical electron velocity in solar wind \citep{Gosling2007} makes $\omega_d = 350000 ~rad/s$.  

\section{Analysis of the Lunar TSI with Magnetized Solar Wind}

The dispersion relation function $D(\omega, k)$ is given by equation (\ref{DisRel}) as

\begin{equation}
D(\omega, k) = 1 - \frac{\omega_b^2}{(\omega - kv_{0b})^2 + \omega_c^2} - \frac{\omega_p^2}{\omega^2}
 \label{DRF}
\end{equation}

\noindent
Now, taking

\begin{equation}
 \frac{n_b}{n_e} = \frac{\omega_b^2}{\omega_p^2} = \alpha
\end{equation}

\noindent
Equation (\ref{DRF}) becomes

\begin{equation}
 D(\omega, k) = 1 - \frac{\alpha\omega_p^2}{(\omega - \omega_d)^2 + \omega_c^2} - \frac{\omega_p^2}{\omega^2}
 \label{DRFm}
\end{equation}

\noindent
To find the local maxima from equation (\ref{DRFm})

\begin{equation}
 \frac{dD(\omega, k)}{d\omega} = 0 = \frac{d}{d\omega}\left[1 - \frac{\alpha\omega_p^2}{(\omega - \omega_d)^2 + \omega_c^2}
 - \frac{\omega_p^2}{\omega^2}\right]
\end{equation}

\noindent
Which becomes

\begin{equation}
 \frac{2\omega_p^2}{\omega^3} + \frac{2\alpha\omega_p^2(\omega - \omega_d)}{[(\omega - \omega_d)^2 + \omega_c^2]^2} = 0
\end{equation}

\noindent
and on solving reduced to

\begin{equation}
 (\omega - \omega_d)^4 + \omega_c^4 + 2(\omega - \omega_d)^2\omega_c^2 + \alpha\omega^3(\omega - \omega_d) = 0
 \label{omega}
\end{equation}

\noindent
For $\alpha << 1$ as $n_b \approx 5 ~cm^{-3}; n_e \approx 300 ~cm^{-3}$, $\omega_d \approx 100\omega_c$ as $v \approx 3.75 \times 10^5 ~m/s; B \approx 5 ~nT$, and neglecting higher order terms, four roots of equation (\ref{omega}) can be obtained as

\begin{eqnarray}
 \omega = \omega_d \pm 2\omega_c - \frac{1}{2}\left(\frac{8}{27}\right)^{3/4} \sqrt{\omega_d\omega_c} \\
 \omega = \omega_d \pm 2\omega_c + \frac{1}{2}\left(\frac{8}{27}\right)^{3/4} \sqrt{\omega_d\omega_c}
\end{eqnarray}

\noindent
For local maxima equation (50) can be taken as

\begin{equation}
\omega = \omega_d \pm 2\omega_c + \frac{1}{2}\left(\frac{8}{27}\right)^{3/4} \sqrt{\omega_d\omega_c} \approx \omega_d + 4\omega_c = \omega_m
\end{equation}

\noindent
The dispersion relation for the local maxima can be obtained from equation (45) as

\begin{equation}
 D_m(\omega_m, k) = 1 - \frac{\alpha\omega_p^2}{(\omega_m - \omega_d)^2 + \omega_c^2} - \frac{\omega_p^2}{\omega_m^2}
\end{equation}

\noindent
Using equation (51) in this, it becomes

\begin{equation}
 D_m(\omega_m, k) = 1 - \frac{17\omega_p^2\omega_c^2 + \alpha\omega_p^2(\omega_d + 4\omega_c)^2}{17\omega_c^2(\omega_d + 4\omega_c)^2}
\end{equation}

\noindent
The local maxima depend upon the beam and plasma parameters. There are two cases here: Either all the roots are real or all the roots are imaginary

\subsection{$D_m > 0$: All roots are real}

Equation (\ref{DisRel}) can be written as

\begin{equation}
 \omega^4 -2 \omega_d \omega^3 + [{\omega_d}^2 + {\omega_c}^2 - Kn_e(1 + \alpha]\omega^2 + [2\omega_dKn_e]\omega - Kn_e(\omega_d^2 + \omega_c^2) = 0
\end{equation}

\noindent
Here, $K = e^2/m\epsilon_0$ and $Kn_e = \omega_p$

\noindent
Now, when the middle term in equation (45) is taken as zero, we are left with

\begin{equation}
1 - \frac{\omega_p^2}{\omega^2} = 0
\end{equation}

\noindent
which gives

\begin{equation}
 \omega = \omega_p = \sqrt{\frac{e^2n_e}{m\epsilon_0}}
\end{equation}

\noindent
And when the last term in equation (45) is taken as zero, we get

\begin{equation}
1 - \frac{\alpha \omega_p^2}{(\omega - \omega_d)^2 +\omega_c^2} = 0
\end{equation}

\noindent
which can be written as

\begin{equation}
 (\omega - \omega_d)^2 +\omega_c^2 = K\alpha n_e
\end{equation}

\noindent
Equation (58) has two roots which are given as

\begin{equation}
 \omega = \omega_d \pm \sqrt{K \alpha n_e - \omega_c^2}
\end{equation}

Now, the phase velocities of the propagating beam modes are given as

\begin{equation}
 v_{\phi} = \frac{\omega}{k} = \frac{\omega_d}{k} \pm \frac{1}{k}\sqrt{\frac{\alpha e^2 n_e}{m\epsilon_0} - \omega_c^2} = v_{0b} \pm \frac{1}{k}\sqrt{\frac{\alpha e^2 n_e}{m\epsilon_0} - \omega_c^2}
\end{equation}

\noindent
and the group velocity of the propagating beam mode is given as

\begin{equation}
 v_g = v_{0b}
\end{equation}

\noindent
The two beam modes are traveling waves with different phase velocity but same group velocity, implies that the transfer of information is being at equal velocity. As group velocity is defined for no. of superimposed waves; so we can say, beam modes are superposition of many waves.

\noindent
Here, we see that group velocity is equal to beam velocity that suggest that beam modes are in resonance with beam itself.

\noindent
The other two modes have zero phase velocity as well as group velocity it’s because plasma oscillation do not carry any information and are oscillating perturbation of plasma parameter.

\noindent
In summary, in this case we see two modes which are due to lunar plasma and other two modes (traveling wave) are only can be seen for high speed beam plasma interaction with lunar plasma.

\subsection{$D_m < 0$: All roots are imaginary}

An instability can get generated if the roots of the dispersion relation are imaginary.

\noindent
From equation (51), $\omega_m$ is inserted in place of $\omega$ in equation (\ref{DRFm}). The resultant equation is

\begin{equation}
 D(\omega, k) = 1 - \frac{\alpha\omega_p^2}{(\omega_d + 4\omega_c^2 - \omega_d)^2 + \omega_c^2} - \frac{\omega_p^2}{\omega_d^2 + 4\omega_c^2} < 0
\end{equation}

\noindent
which becomes

\begin{equation}
 \frac{\alpha\omega_p^2}{17\omega_c^2} + \frac{\omega_p^2}{\omega_d^2 + 4\omega_c^2} > 1
\end{equation}

\noindent
and gives

\begin{equation}
 \omega_d < -4\omega_c \pm \sqrt{\frac{17\omega_c^2\omega_p^2}{17\omega_c^2 - \alpha\omega_p^2}}
\end{equation}

\noindent
The condition for instability to occur is obtained by replacing the terms $\omega_d$, $\omega_c$, $\omega_p$ and $\alpha$ and it comes out to be

\begin{equation}
 k < - \frac{4eB_0}{mv_{0b}} \pm \frac{1}{v_{0b}}\sqrt{\frac{17e^2B_0^2n_{0e}}{m(17B_0^2 \epsilon_0 - n_{0b})}}
 \label{TSIcon}
\end{equation}

\noindent
Equation (\ref{TSIcon}) shows that the instability depends upon the lunar electron density ($n_{0e}$), the electron beam velocity ($v_{0b}$), the electron beam density ($n_{0b}$), and the magnetic field ($B_0$).

\subsection{Growth Factor of the TSI}

When local maxima goes below zero, out of the four roots, two roots vanishes and only the roots giving the plasma and beam modes exists.

\noindent
The $D(\omega, k)$ is expanded about the local maxima which is given as

\begin{equation}
 D (\omega) = D (\omega_m) + (\omega - \omega_m)\frac{dD}{d\omega} + \frac{1}{2!}(\omega - \omega_m)^2\frac{d^2D}{d\omega^2} + ...
\end{equation}

\noindent
Now, $D'(\omega) = 0$, and for local maxima $D''(\omega_m) < 0$.

\noindent
The condition for dispersion relation is given as

\begin{equation}
 D(\omega) = 0 = D (\omega_m) + (\omega - \omega_m)\frac{dD}{d\omega} + \frac{1}{2!}(\omega - \omega_m)^2\frac{d^2D}{d\omega^2} + ...
\end{equation}

\noindent
Equation (67) is a quadratic and is solved for $(\omega - \omega_m)$ to get

\begin{equation}
 (\omega - \omega_m) = \frac{-D'(\omega) \pm \sqrt{[D'(\omega_m)]^2 - 2D''(\omega_m)D(\omega_m)}}{D''(\omega_m)}
\end{equation}

\noindent
For $D'(\omega) = 0$, this equation reduces to

\begin{equation}
 (\omega - \omega_m) = \pm \frac{\sqrt{- 2D''(\omega_m)D(\omega_m)}}{D''(\omega_m)}
\end{equation}

\noindent
which gives

\begin{equation}
 \omega = \omega_m \pm i\sqrt{\frac{2D(\omega_m)}{D''(\omega_m)}}
\end{equation}

\noindent
The real part of this complex root is given as

\begin{equation}
 \omega_{re} = \omega_m = \omega_d + 2\omega_c + \frac{1}{2}\left(\frac{8}{27}\right)^{3/4} \sqrt{\omega_d\omega_c}
\end{equation}

\noindent
The phase velocity of the instability is due to its real part which is now given as

\begin{equation}
 v_{\phi} = \frac{\omega_{re}}{k} = \frac{\omega_m}{k} = \frac{1}{k}\left[\omega_d + 2\omega_c + \frac{1}{2}\left(\frac{8}{27}\right)^{3/4}\sqrt{\omega_d\omega_c} \right]
\end{equation}

\noindent
The group velocity of the instability is given as

\begin{equation}
 v_g = \frac{d\omega_{re}}{dk} = v_{\phi}
\end{equation}

\noindent
The imaginary part of the instability give the growth rate as given by

\begin{equation}
 \omega_{im} = \sqrt{\frac{2D(\omega_m)}{D''(\omega_m)}}
\end{equation}

\noindent
$D''(\omega_m)$ is obtained from equation (\ref{DRFm}) as

\begin{equation}
 D''(\omega) = - \frac{6\omega_p^2}{\omega^4} - \frac{\omega_b^2[6(\omega - \omega_d)^2 - 2\omega_c^2]}{[(\omega - \omega_d)^2 + \omega_c^2]^3}
\end{equation}

\noindent
For $\omega = \omega_m$, this equation becomes

\begin{equation}
 D''(\omega_m) = - \frac{6\omega_p^2}{\omega_m^4} - \frac{\omega_b^2[6(\omega_m - \omega_d)^2 - 2\omega_c^2]}{[(\omega_m - \omega_d)^2 + \omega_c^2]^3}
\end{equation}

\noindent
or it can be written as

\begin{equation}
 D''(\omega_m) = - \frac{6\times17^3\omega_p^2\omega_c^4 + 94\alpha \omega_p^2(\omega_d + 4\omega_c)^4}{17^3\omega_c^4(\omega_d + 4\omega_c)^4}
\end{equation}

\noindent
By inserting equations (\ref{DRFm}) and equation (77) in equation (74), $\omega_{im}/\omega_p$ is obtained as

\begin{equation}
 \frac{\omega_{im}}{\omega_p} = 17\omega_c\left(\frac{\omega_d}{\omega_p} + \frac{4\omega_c}{\omega_p}\right)\sqrt{\frac{\alpha \omega_p^2\left(\frac{\omega_d}{\omega_p} + \frac{4\omega_c}{\omega_p}\right)^2 + 17\omega_c^2\left[1 - \left(\frac{\omega_d}{\omega_p} + \frac{4\omega_c}{\omega_p}\right)^2\right]}{29478\omega_c^4 + 94\alpha\omega_p^4\left(\frac{\omega_d}{\omega_p} + \frac{4\omega_c}{\omega_p}\right)^4 }}
 \label{Growthfac}
\end{equation}

\noindent
It can be seen from equation (\ref{Growthfac}) that the growth factor depend on the plasma, beam parameter and wave vector $k$ values of the wave. Depending on the value of $k$, there are going to be different instabilities getting generated at certain altitude of lunar ionosphere. The growth rate for various values of $n$ is compared with $k$ to explore the fastest growing mode.

\noindent

The plasma parameter values taken for the estimation of the growth factor is as follows: The normal solar wind parameters, the solar wind electron number density, ($n_b$) is taken as $5 ~cm^{-3}$ and the solar wind electron velocity, ($v_{0b}$) is taken as $3.75 \times 10^5 ~m/s$ \citep{Gosling2007}. The lunar plasma number density in lunar ionosphere at various altitudes is estimated to be typically in the range $10 - 300 ~cm^{-3}$ from the radio occultation (RO) measurements \citep{Choudhary2016}. The typical solar wind magnetic field, $B_0$ is taken as $5 ~nT$ \citep{Gosling2007}. 

It is to be noted that the solar wind and lunar plasma parameters taken in this study are same as that of in earlier studies \citep{Chakraborty2023} and \citep{Yadav2024b} for the continuity and comparison.

\begin{figure}
\centering
\includegraphics[height=2.10in,width=3.25in]{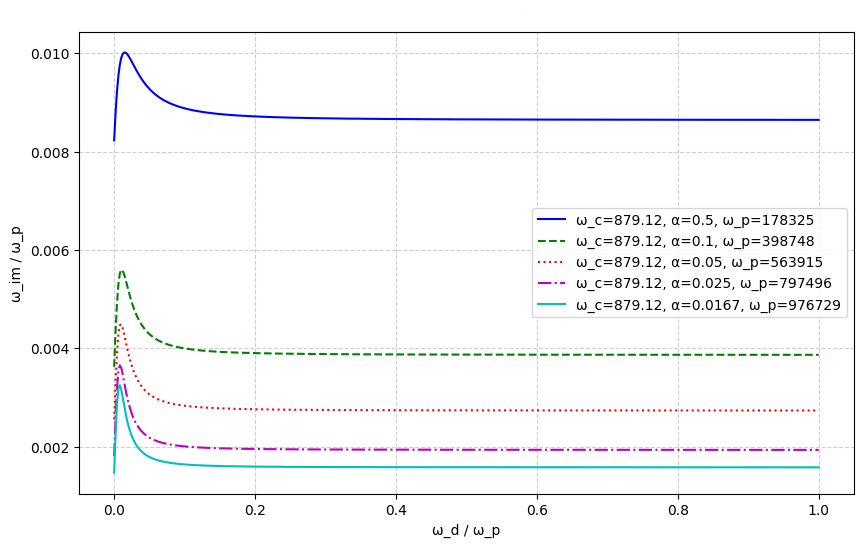}\\ \vspace{2mm}
\includegraphics[height=2.10in,width=3.25in]{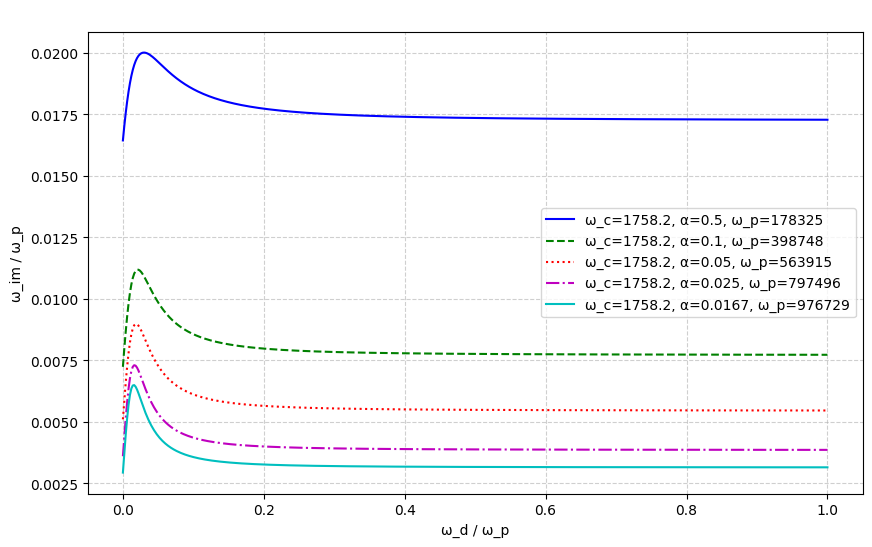}\\ \vspace{2mm}
\includegraphics[height=2.10in,width=3.25in]{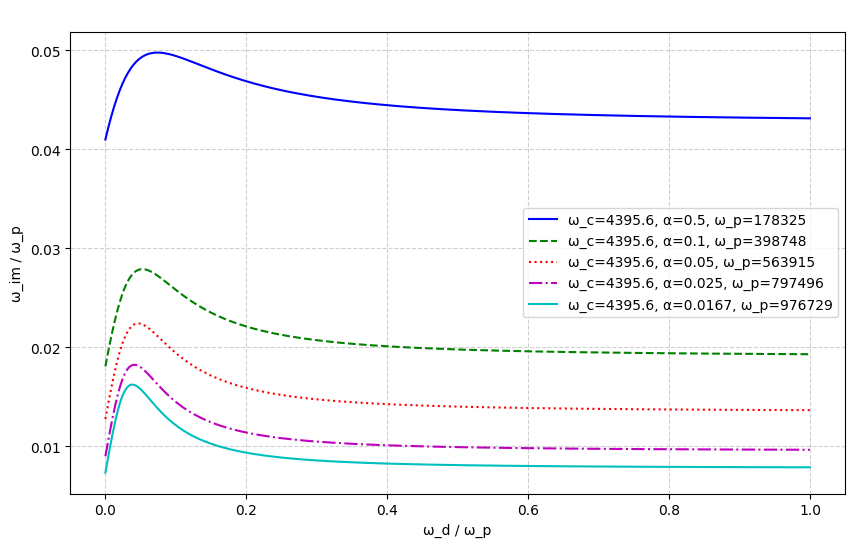}\\ \vspace{2mm}
\includegraphics[height=2.10in,width=3.25in]{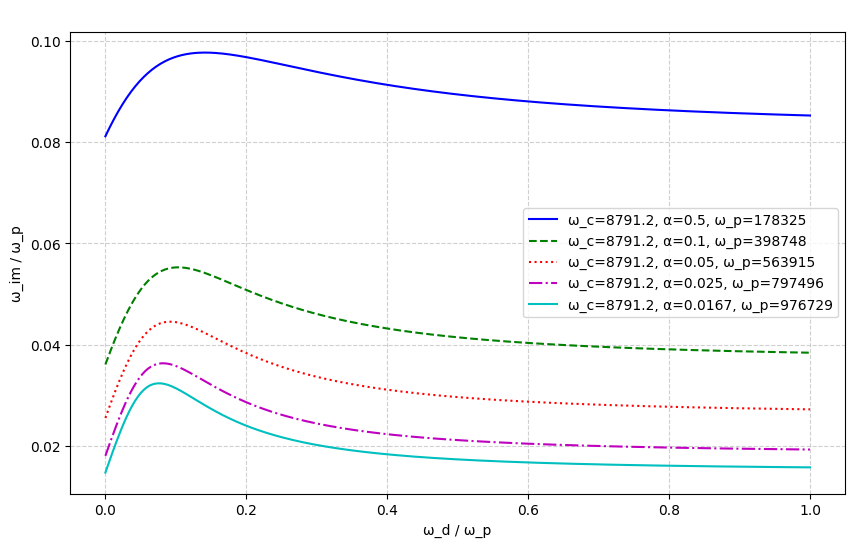}
\caption{The growth rate when: $B_0 = 5 ~nT$ (top); $B_0 = 10 ~nT$ (second from top); $B_0 = 25 ~nT$ (third from top); $B_0 = 50 ~nT$ (bottom).}
\label{GRB}
\end{figure}

With the above mentioned plasma and magnetic field parameters, the $k$ as a ratio of $\omega_d/\omega_p$ is plotted with the instability growth rate ($\omega_{im})/\omega_p$ for a preset IMF ($B_0$), variable $\alpha$ and various plasma densities as shown in Figures (\ref{GRB}) where $B_0$ is taken between $5 - 50 ~nT$. The solar wind magnetic field can have these higher magnitudes of $B_0 = 25 - 50 ~nT$ or more when there is a solar magnetic storm hitting the Moon during which the other solar wind parameters can remain unchanged.

\noindent
Figure \ref{GRB} (top) shows that for $B_0 = 5 ~nT$, the growth rate initially increases very fast; then drops down as the ratio $\omega_d/\omega_p$ increases and becomes almost constant for the higher values of the ratio. A similar behavior is observed for $B_0 = 10 ~nT$ shown in Figure \ref{GRB} (second from top), however, the fall is not as steep as the earlier case but also becomes constant for this magnetic field. For $B_0 = 25 ~nT$, Figure \ref{GRB} (second from bottom) and for $B_0 = 50 ~nT$, Figure \ref{GRB} (bottom) this trend of broadening of the peak continued with increasing $\omega_d/\omega_p$ ratio however, both the rise and the fall of the growth rate has got smoothen. Also, for higher magnetic fields, the growth rate slope which is observed to be constant for the lower IMF magnitudes, now tends to clearly decline with the increase in ratio. It is to be noted that the growth rate is highest for the lowest lunar electron density of $n_e = 10 ~cm^{-3}$ which is obvious due to the fact that with this smallest electron density, both the growth rate and the wave vector are normalized which led to the increase in the magnitude.

\noindent
The parameters and the ranges used to obtain the growth rate with variable IMF magnitudes are summarized in table \ref{table1}.

\begin{table}[]
\caption{\label{table1}Instability growth rate for different IMF magnitude.}
\begin{tabular}{llllllll}
\hline
\multicolumn{2}{c}{B = 5 nT}    & \multicolumn{2}{c}{B = 10 nT}   & \multicolumn{2}{c}{B = 25 nT}   & \multicolumn{2}{c}{B = 50 nT}   \\
\hline
\multicolumn{2}{l}{$\omega_c$ = 879.12} & \multicolumn{2}{l}{$\omega_c$ = 1758.2} & \multicolumn{2}{l}{$\omega_c$ = 4395.6} & \multicolumn{2}{l}{$\omega_c$ = 8791.2} \\
\hline
$\alpha$ & $\omega_p$ & $\alpha$ & $\omega_p$ & $\alpha$ & $\omega_p$ & $\alpha$ & $\omega_p$ \\
0.5     & 178325     & 0.5    & 178325  & 0.5    & 178325  & 0.5     & 178325  \\
0.1     & 398748     & 0.1    & 398748  & 0.1    & 398748  & 0.1     & 398748  \\
0.05    & 563915     & 0.05   & 563915  & 0.05   & 563915  & 0.05    & 563915  \\
0.025   & 797496     & 0.025  & 797496  & 0.025  & 797496  & 0.025   & 797496  \\
0.0167  & 976729     & 0.0167 & 976729  & 0.0167 & 976729  & 0.0167  & 976729 \\ 
\hline\end{tabular}
\end{table}

\noindent
In the next step, in order to find the fastest growing mode, $d\omega_{im}/d\omega_d = 0$. Hence, on differentiating equation (\ref{Growthfac}) with respect to $\omega_d$ gives

\begin{equation}
 \alpha \omega_p^2(\omega_d + 4\omega_c)^4 - 36\omega_c^2(\alpha \omega_p^2 - 17\omega_c^2)(\omega_d + 4\omega_c)^2 - 306\omega_p^2\omega_c^4 = 0
\end{equation}

\noindent
This equation is solved and the condition for fastest growth rate is obtained as

\begin{equation}
 k = - \frac{4\omega_c}{v_{0b}} \pm \frac{1}{v_{0b}}\sqrt{\frac{36\omega_c^2(\alpha \omega_p^2 - 17\omega_c^2) \pm \sqrt{1296\omega_c^2(\alpha \omega_p^2 - 17\omega_c^2)^2 + 1224\alpha \omega_p^4 \omega_c^4}}{2 \alpha \omega_p^2}}
\end{equation}

\begin{figure}
\centering
\includegraphics[height=2.10in,width=3.25in]{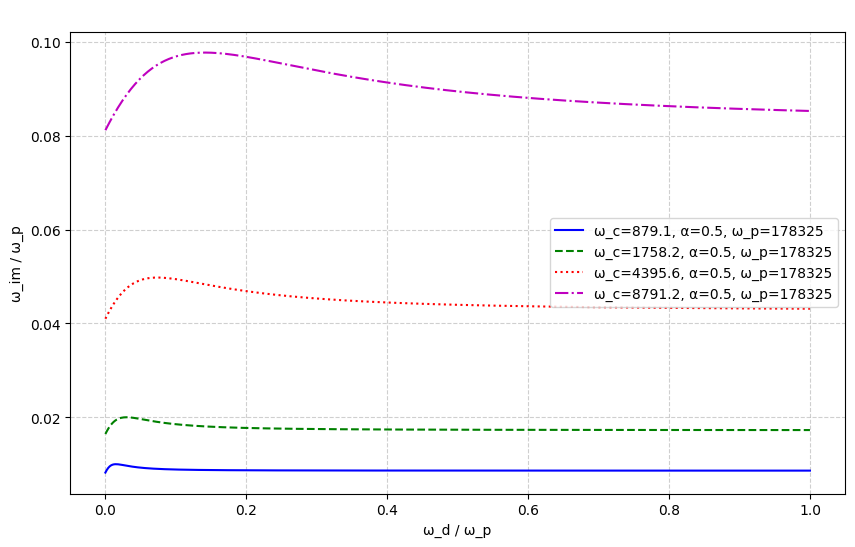}\\
\includegraphics[height=2.10in,width=3.25in]{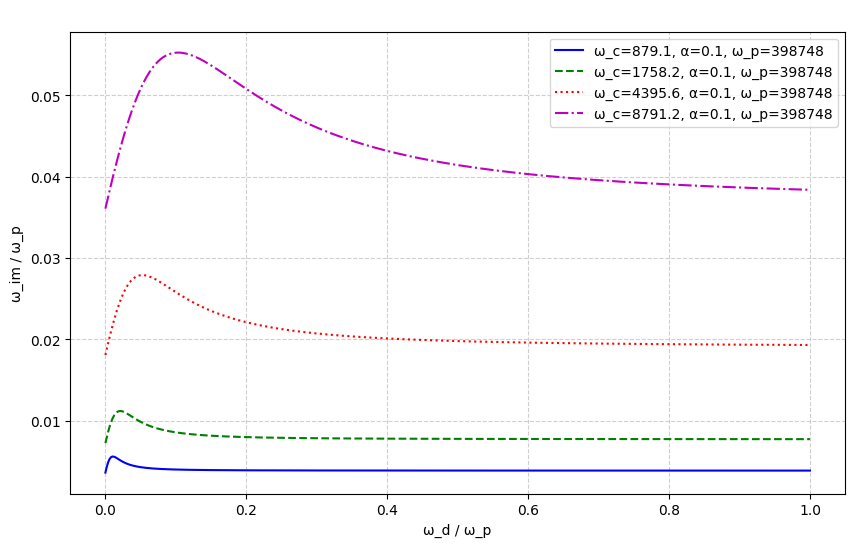}
\caption{The growth rate for: $n_e = 10 ~cm^{-3}$ (top); $n_e = 50 ~cm^{-3}$ (bottom).}
\label{GRN1}
\end{figure}

\noindent
Further, the $k$ as a ratio of $\omega_d/\omega_p$ is plotted with the instability growth rate ($\omega_{im})/\omega_p$ for a preset lunar electron plasma density ($n_e$) of $10$ and $50 ~cm^{-3}$, various IMF magnitudes, $B_0 = 5, 10, 25, 50 ~ nT$ respectively and a variable $\alpha = n_b/n_e$ for $n_b = 5 ~cm^{-3}$ as shown in Figure \ref{GRN1}.

\begin{figure}
\centering
\includegraphics[height=2.10in,width=3.25in]{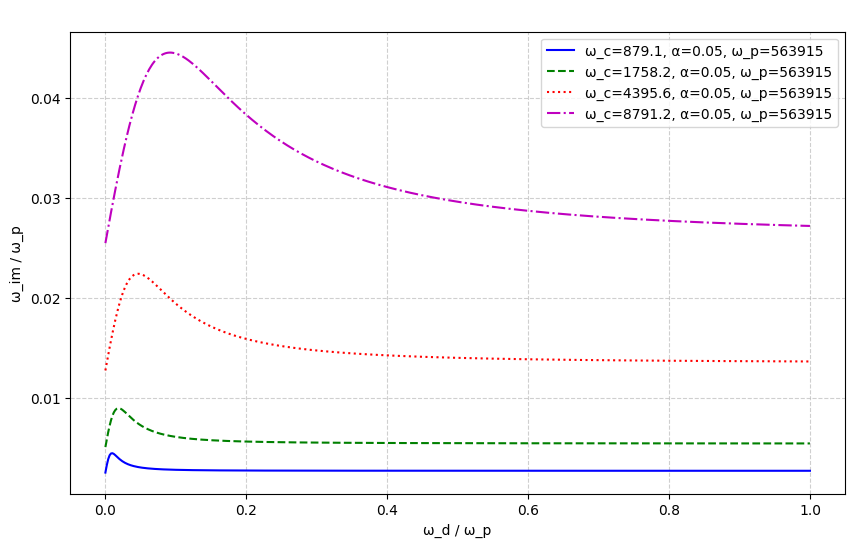}\\
\includegraphics[height=2.10in,width=3.25in]{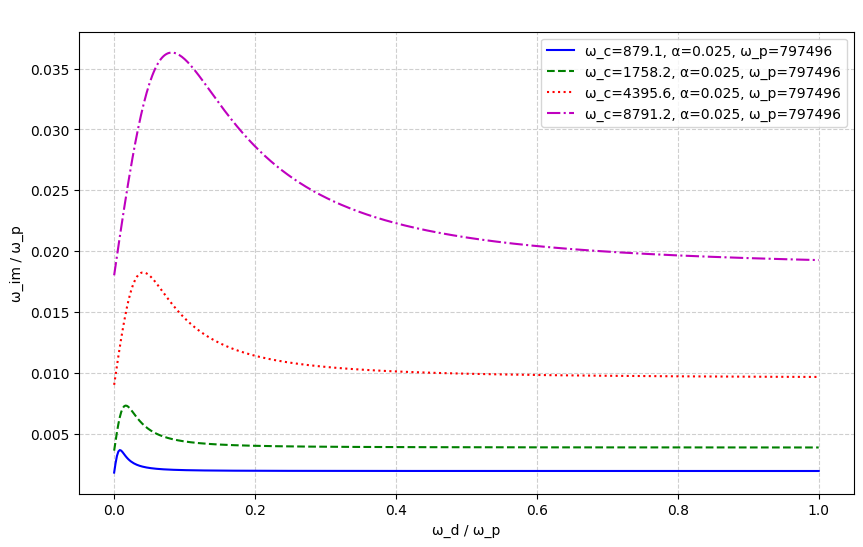}\\
\includegraphics[height=2.10in,width=3.25in]{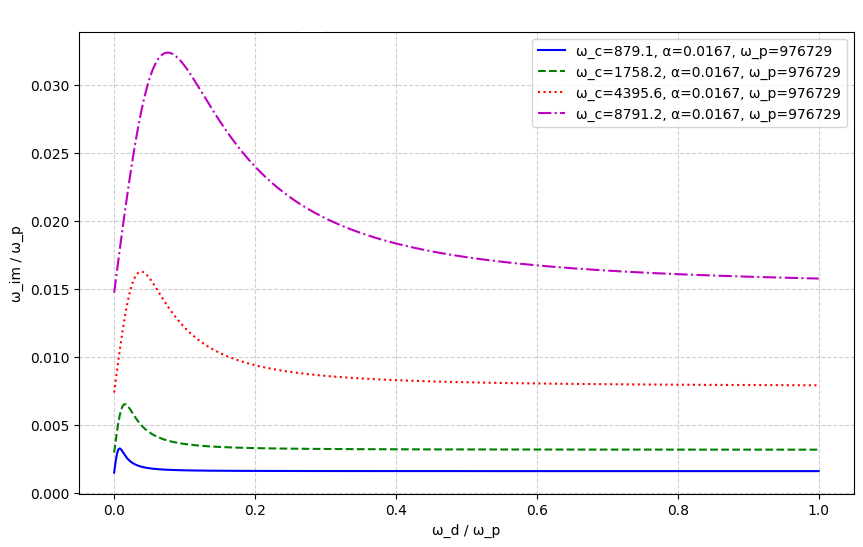}
\caption{The growth rate for: $n_e = 100 ~cm^{-3}$ (top); $n_e = 200 ~cm^{-3}$ (middle); and $n_e = 300 ~cm^{-3}$ (bottom).}
\label{GRN2}
\end{figure}

\noindent
And for the lunar electron density ($n_e$) in the range [$100, 200, 300 ~cm^{-3}$], it is shown in Figure \ref{GRN2}.

\noindent
In Figure \ref{GRN1}, (top), $n_e$ is taken as $10 ~cm^{-3}$ and $\alpha = 0.5$. The growth factor initially increases slowly and then slowly declines and become almost constant with increase in $k$. The maximum growth factor is observed for the highest IMF magnitude $B_0 = 50 ~ nT$ showing its influence of IMF on the lunar TSI. 

Figure \ref{GRN1}, (bottom) shows the trend for $n_e = 50 ~cm^{-3}$ with varied IMF in the preset range and now a changed $\alpha = 0.1$ due to a change in $n_e$. As can be seen with increased lunar electron density, the rise and the fall of the growth rate becomes steep and the peak shrinks as the ratio $\omega_d/\omega_p$ is increased.   

For $n_e = 100 ~cm^{-3}$, the $\alpha$ drops further to become $\alpha = 0.05$ as shown in Figure \ref{GRN2}, (top) for all the IMF magnitudes. With an increase in $k$, the instability growth factor peak shrinks for all the IMF magnitudes making the rise and fall more sharper. 

In Figure \ref{GRN2}, (middle), the growth rate for $n_e = 200 ~cm^{-3}$ when $\alpha = 0.025$ further drops down due to the higher lunar electron density. For all the IMF magnitudes, the growth factor rise and fall becomes more steeper than the previous cases and the peak for maximum growth becomes even more sharper.     

Figure \ref{GRN2}, (bottom) shows the instability growth factor for the 
highest lunar electron number density of $n_e = 300 ~cm^{-3}$, for which $\alpha$ becomes $= 0.0167$ and all the IMF magnitudes. As can be seen from the figure that the rise of the growth factor is very steep and the fall has also become steeper than the earlier cases. For all IMF magnitudes, the plot further shrinks. For lower IMF magnitudes, the growth rate becomes absolutely flat after the initial hike.      

\noindent
The parameters and the ranges used to obtain the growth rate with variable lunar electron plasma density are summarized in table \ref{table2}.

\begin{table}
\caption{\label{table2}Instability growth rate for different lunar plasma densities.}
\begin{tabular}{lccccr}
\hline
$n_e = 10 ~cm^{-3}$ & $n_e = 50 ~cm^{-3}$ & $n_e = 100 ~cm^{-3}$ & $n_e = 200 ~cm^{-3}$ & $n_e = 300 ~cm^{-3}$ \\
\hline
$\omega_p = 178325$ & $\omega_p = 398748$ & $\omega_p = 563915$ & $\omega_p = 797496$ & $\omega_p = 976729$ \\
\hline
$\alpha$ = 0.5 & $\alpha$ = 0.1 & $\alpha$ = 0.05 & $\alpha$ = 0.025 & $\alpha$ = 0.0167 \\
\hline
$\omega_c$ & $\omega_c$ & $\omega_c$ & $\omega_c$ & $\omega_c$ \\
879.2      & 879.2      & 879.2      & 879.2      & 879.2      \\
1758.2     & 1758.2     & 1758.2     & 1758.2     & 1758.2     \\ 
4395.6     & 4395.6     & 4395.6     & 4395.6     & 4395.6     \\
8791.2     & 8791.2     & 8791.2     & 8791.2     & 8791.2     \\
\hline
\end{tabular}
\end{table}

\section{Solar Wind - Lunar Plasma Interaction: PIC Simulations} 

The solar wind - lunar plasma interaction is visualized using the particle-in-cell (PIC) simulations. The conditions for PIC simulations are exactly similar to those adopted in earlier works [\cite{Chakraborty2023} and \cite{Yadav2024b}]. It is considered here that the lunar plasma is ``cold'' and that no energetic electrons are present so that the thermal effects can be neglected. Furthermore, in these simulations only the solar wind is considered as magnetized and the lunar plasma is taken as non-magnetized.

\noindent
A total of $40,000$ particles are taken for this PIC simulation along with 390 number of mesh cells. A box size of 50 is taken in the periodic domain. The PIC simulation outcomes are in the form of phase diagrams where the position is taken on the x-axis and the velocity is taken on y-axis. It is to be noted that in the phase diagram, the red line represent the lunar electron background whereas the blue dots indicate the incoming solar wind. During these simulations a time step of $0.1 ~s$ is taken. 

Moreover, the solar wind - lunar plasma interactions for four different magnitudes of the IMF namely $B_0 = 5, 10, 25, 50 ~nT$ are simulated here. The solar wind velocity, $v_{0b}$ is taken as $3.5 \times 10^5 ~m/s$. 

\begin{figure}
\centerline{\hspace*{0.5in}
\includegraphics[height=8.0in,width=6.0in]{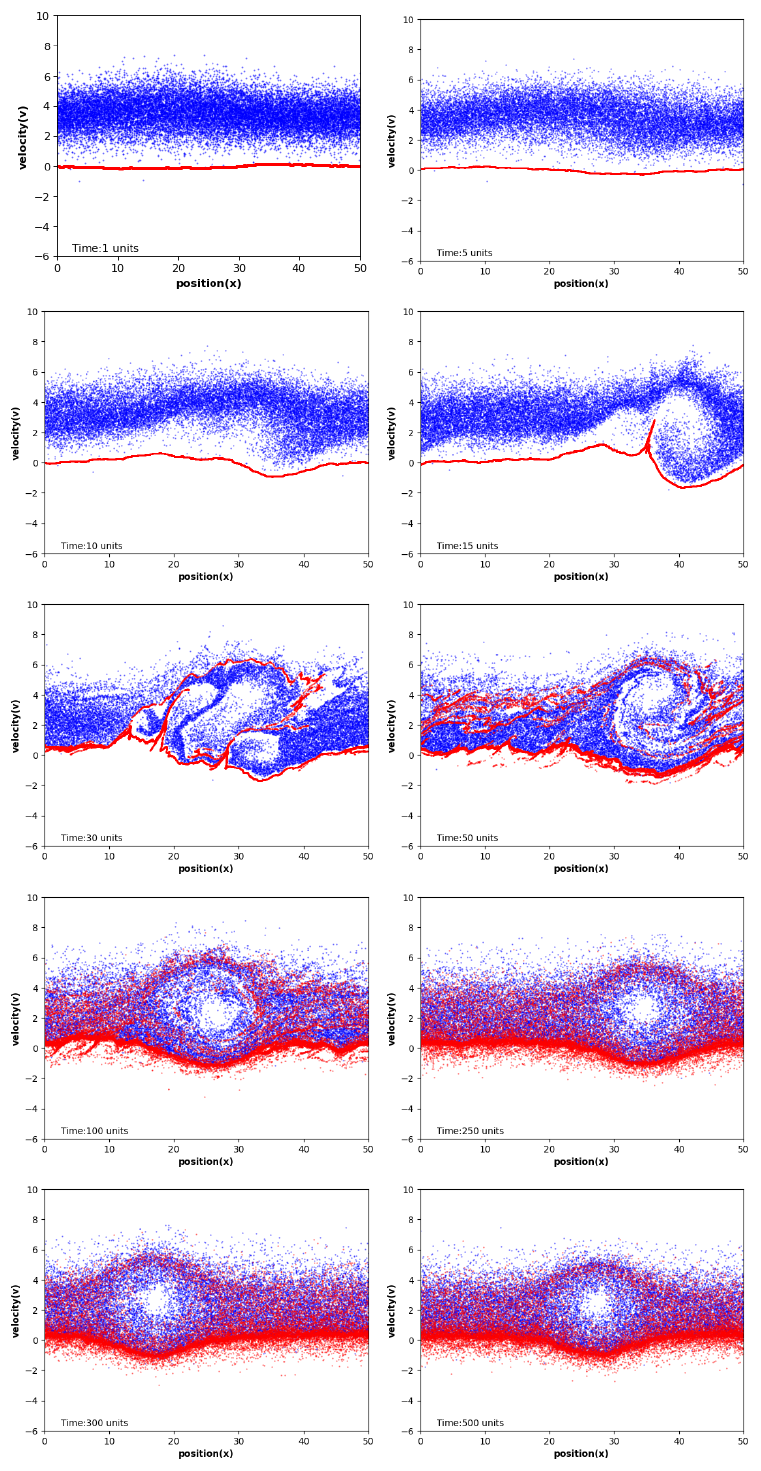}}
\caption{The PIC simulations for the lunar TSI when $B_0 = 5 ~nT$ is taken as the IMF for t = 1 to 500 units. Red line represents the lunar plasma and the solar wind plasma is in blue color.}
\label{B05nt}
\end{figure}

Figures \ref{B05nt} shows the phase diagrams for the normal solar wind condition with $B_0 = 5 ~nT$ and the time steps: t = 1, 10, 15, 30, 50, 100, 250, 300, and 500 units. 

\noindent
Figure \ref{B05nt} shows that the mixing of solar wind electrons with lunar electrons started around t = 30 units and the vortex starts forming after t = 50 units. After t = 100 units the vortex is formed and at t = 500 units, the vortex shape is retained but the whole plasma ensemble appears slightly squeezed. 

\begin{figure}
\centerline{\hspace*{0.5in}
\includegraphics[height=8.0in,width=6.0in]{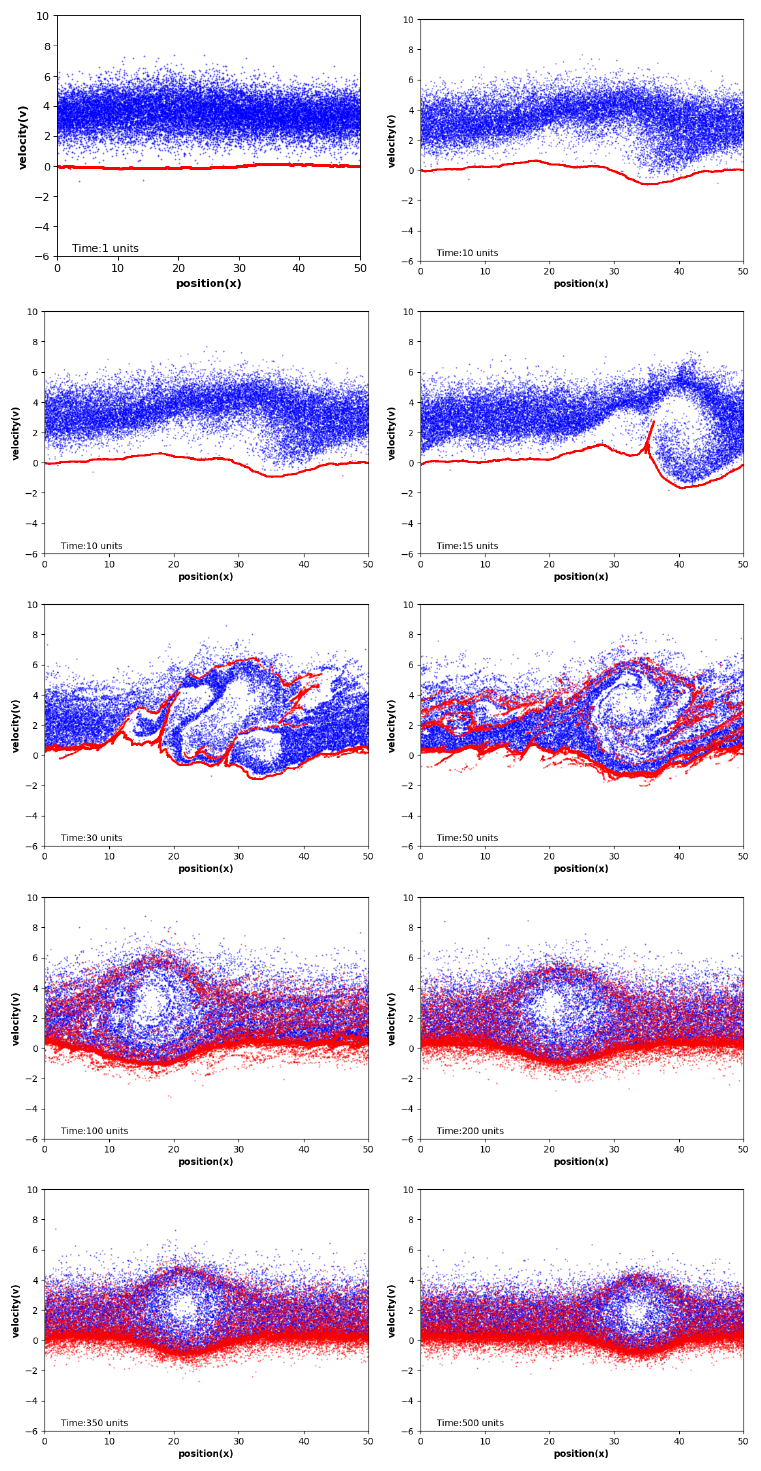}}
\caption{The PIC simulations for the lunar TSI when $B_0 = 10 ~nT$ is taken as the IMF for t = 1 to 500 units. Red line represents the lunar plasma and the solar wind plasma is in blue color.}
\label{B10nt}
\end{figure}

Figures \ref{B10nt} shows the phase diagrams for the normal solar wind condition with $B_0 = 10 ~nT$ and the time steps: t = 1, 10, 15, 30, 50, 100, 200, 350, and 500 units. 

As can be seen from Figure \ref{B10nt}, the solar wind electrons interact with the lunar electrons now for a slightly raised IMF magnitude with $B_0 = 10 ~nT$. Here, also the phase diagrams for 9 time steps are shown. For this IMF magnitude also, the vortex formation is observed along with a new phenomena of shielding of the solar wind electrons by the lunar electrons. The shrinking of the interacting plasma populations continues.

\begin{figure}
\centerline{\hspace*{0.5in}
\includegraphics[height=8.0in,width=6.0in]{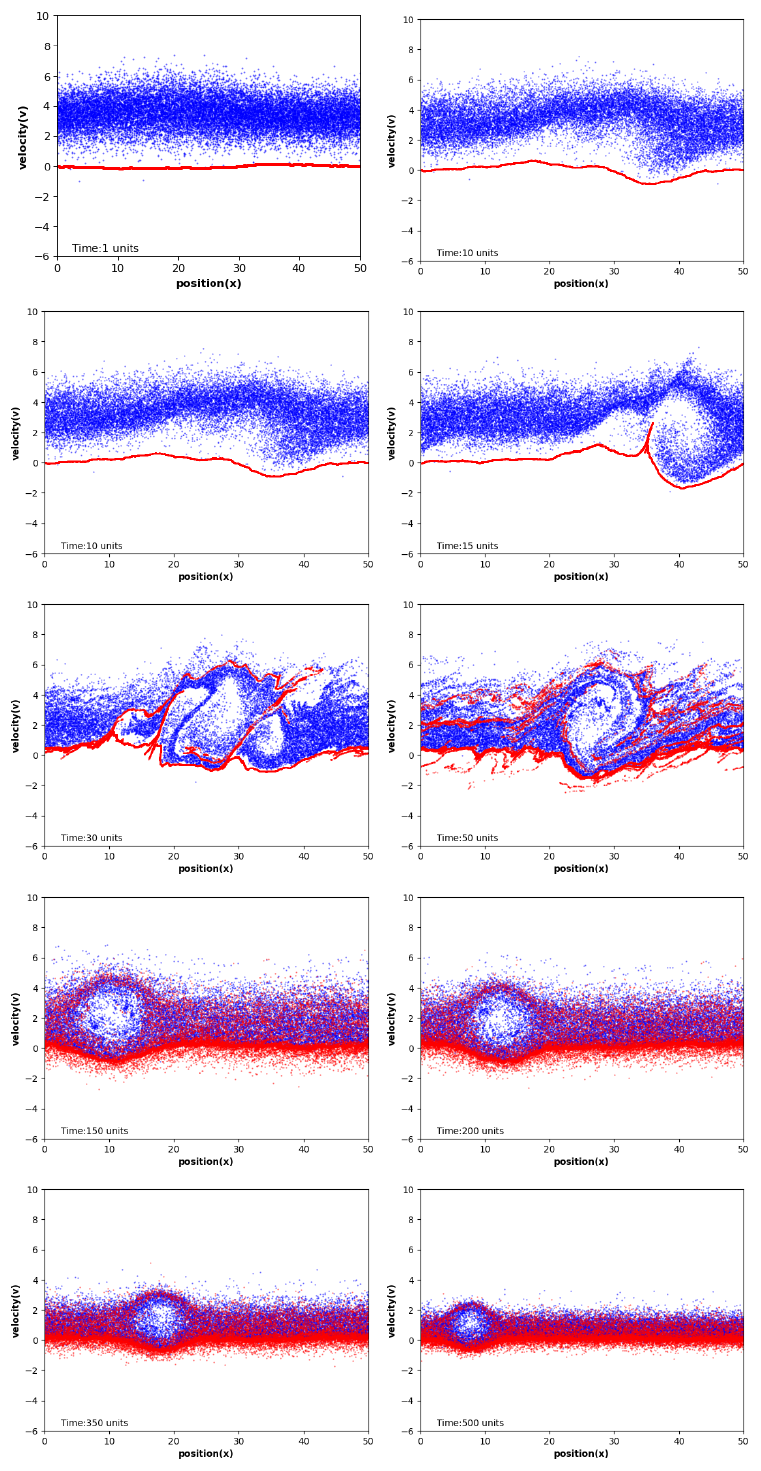}}
\caption{The PIC simulations for the lunar TSI when $B_0 = 25 ~nT$ is taken as the IMF for t = 1 to 500 units. Red line represents the lunar plasma and the solar wind plasma is in blue color.}
\label{B25nt}
\end{figure}

Figures \ref{B25nt} shows the phase diagrams for the normal solar wind condition with $B_0 = 25 ~nT$ and the time steps: t = 1, 10, 15, 30, 50, 150, 200, 350, and 500 units. 

\noindent
Figure \ref{B25nt} depicts the phase diagrams for 9 time steps of the solar wind - lunar plasma interactions for a possible condition when there is a solar magnetic storm that is hitting the lunar plasma. As can be seen from the figure, a smooth vortex is formed by the time step t = 100. The shielding of the solar wind electrons by the lunar electrons as well as the shrinking of the interaction region is unambiguously visible now for a higher IMF magnitude.

\begin{figure}
\centerline{\hspace*{0.5in}
\includegraphics[height=8.0in,width=6.0in]{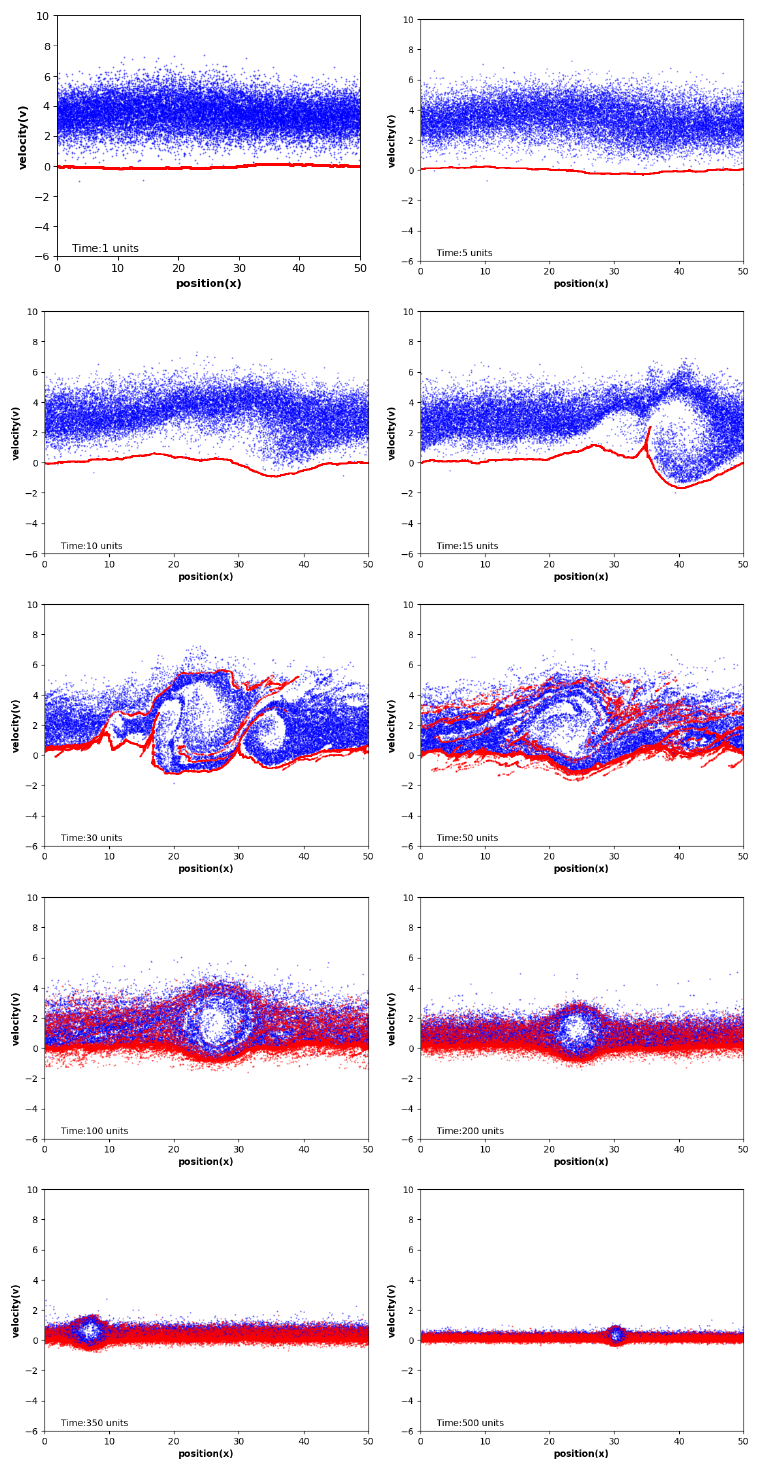}}
\caption{The PIC simulations for the lunar TSI when $B_0 = 50 ~nT$ is taken as the IMF for t = 1 to 500 units. Red line represents the lunar plasma and the solar wind plasma is in blue color.}
\label{B50nt}
\end{figure}

Figures \ref{B50nt} shows the phase diagrams for the normal solar wind condition with $B_0 = 50 ~nT$ and the time steps: t = 1, 10, 15, 30, 50, 100, 200, 350, and 500 units respectively. 

The interaction behavior of magnetized solar wind - lunar electron plasma in Figure \ref{B50nt} is also similar to the earlier figure of higher magnitude and the same inferences can be drawn here. The vortex formation and the interaction region getting restricted are visible. However, a new phenomena now appears which shows that only the lunar electrons are visible now (red color) which means either all the solar wind electrons are either completely shielded or the solar wind electrons after interacting with the lunar electrons have become the lunar electrons itself after delineating from the IMF!!! This issue needs more studies for a suitable clarification.

\section{Discussion}

The growth rate of the TSI in lunar plasma environment is estimated which depends upon the solar wind parameters, the IMF magnitude and the lunar electron plasma density. The maximum growth rate is also estimated with the combinations of various lunar electron number density with different magnitudes of the IMF. The comparison of maximum growth rate for the streaming instability normalized with angular lunar electron plasma frequency with normalized $\omega_d$ is carried out. 

Initially, for a fixed IMF magnitude, the fastest growth rate is plotted for a number of lunar electron number densities in the range $n_e = 10 - 300 ~cm^{-3}$. The results point out that as IMF magnitude is increased, the fastest growth rate ratio $\omega_{im})/\omega_p$ also increases from 0.01 for $B_0 = 5 ~nT$ to 0.1 for $B_0 = 50 ~nT$. Also, the the growth rate initially increases very fast and then drops down as the ratio $\omega_d/\omega_p$ increases and becomes almost constant for the higher values of the ratio. For the lower IMF values the growth peak is narrow which expends for higher IMF numbers.  

Afterwards, for a fixed lunar electron number density (from 10 to 300 $~cm^{-3}$) , the fastest growth rate is plotted with $\omega_d/\omega_p$ for IMF magnitudes such as $B_0 = 5, 10, 25, 50 ~ nT$. The results shows that as the lunar number density is increased, the maximum growth rate ratio decreases from 0.1 for $n_e = 10 ~cm^{-3}$ to 0.032 for $n_e = 300 ~cm^{-3}$. This observation is exactly opposite to the previous outcome. Also, contrary to the earlier observation, in this study the growth rate initially increases very slowly and then becomes steep as the ratio $\omega_d/\omega_p$ increases and later falls slowly for the higher values of the ratio. For the lower lunar number densities, the growth peak is broad which shrinks for higher lunar electron number densities. 

The magnetized solar wind - lunar electron plasma interaction is visualized with PIC simulations. Four such simulational runs are carried out for four IMF magnitudes of $B_0 = 5, 10, 25, 50 ~ nT$. The PIC simulations clearly shows that the solar wind - lunar plasma interaction get completed by t = 100 units when the vortex forms. However, keeping the simulations on beyond t = 100 units results in the shrinking of the plasma interaction region. In the vortices, the lunar electrons seem to shield the lunar electrons but in the surrounding arm of the vortex, both the electron populations get scrambled. The simulations also reveals that on increasing the IMF magnitude in the solar wind, the movement of both the solar wind as well as the lunar electrons becomes restricted. Intriguingly, for the maximum IMF strength of $B_0 = 50 ~ nT$ shows the complete dominance of the lunar electrons in the interaction which needs further investigation.   

In this limited study only the non-energetic “cold” lunar electrons are considered and the energetic “hot” lunar are not taken into account. However, in actual scenarios, there can be a smaller population of energetic lunar electrons which may affect the solar wind - lunar plasma interaction. Here, the ion dynamics is avoided primarily due to the low ion density in the lunar ionosphere as compared to the electrons. The effect of lunar crustal magnetic field is also not accounted for in this analysis which can affect the electron dynamics if the magnetized solar wind interaction is taking place near a mini-magnetosphere on the surface of Moon. However, in further studies to be carried out in the near future, all these aspects shall be included in the analysis most likely one at a time in order to clearly mark the effect due to that specific parameter.

\section{Conclusion} 

The magnetized solar wind - lunar electrons interaction is studied analytically during varied plasma conditions which gives rise to the two stream-instability (TSI) generation there. Only the non-energetic lunar electrons are considered to participate in this interaction. In this study, the modified dispersion relation of the TSI is obtained analytically taking into account the interplanetary magnetic field (IMF) also. The growth rate of this streaming instability is derived from the dispersion relation and the maximum (fastest) growth rate is also obtained. The analysis indicate that apart from the solar wind electron - velocity and number density, the IMF also plays a part along with lunar electron number density in the growth of this instability. 

The solar wind - lunar plasma interaction is treated numerically also through the PIC simulations which show that during the solar wind - moon plasma interaction, the local lunar electrons make a shield around the solar wind electrons but only in the vortices and in the surrounding arm of the vortex, both the electrons get scrambled. It is also observed that as the IMF magnitude is increased, the combined population of solar wind and lunar electrons gets squeezed thereby restricting their motion in both the directions. For the maximum IMF magnitude, the interaction appears to be completely dominated by the lunar electrons.       

\begin{acks}
The authors thank Dr. R.K. Choudhary (Head, RSIM), SPL, VSSC for his useful suggestions during the compilation of this manuscript.  
\end{acks}

 
\begin{authorcontribution}
\vspace{2mm}
\noindent
VKY: Original Draft / Visualization / Validation / Conceptualization / Review and editing; AS: Validation / Investigation / Formal analysis / Review and editing; RK: Supervision / Review and editing
\end{authorcontribution}

\begin{conflict}
The authors declare that they have no conflicts of interest.
\end{conflict}

\vspace{5mm}

\noindent
{\bf Ethics}: Ethics declaration: Not applicable.\\

\noindent
{\bf Funding}: There is no funding for this work.
  

\end{document}